\documentclass[
               reprint, twocolumn,
               preprintnumbers,
               amsmath,
               amssymb,
               showpacs,
               prl,
               superscriptaddress]{revtex4-1}

 \usepackage{ucs}
 \usepackage[utf8x]{inputenc}
 \usepackage{latexsym}
 \usepackage{amsfonts}
 \usepackage{amsmath}
 \usepackage[dvips]{graphicx}
 \usepackage{color}
 \usepackage{bm}
 \usepackage{hyperref}
\hypersetup{
   bookmarks=true,         
   colorlinks=true,        
   linkcolor=magenta,      
   citecolor=magenta,      
   filecolor=magenta,      
   urlcolor=blue           
}

\begin{document}

\author{Oriol Vendrell}
\email[e-mail: ]{oriol.vendrell@pci.uni-heidelberg.de}
\affiliation{
    Department of Physics and Astronomy,
    Aarhus University,
    Ny Munkegade 120, 8000 Aarhus C, Denmark}
\altaffiliation[Current affiliation: ]{
    Theoretische Chemie,
    Physikalisch-Chemisches Institut,
    Universität Heidelberg,
    INF 229, 69120 Heidelberg, Germany}


\title{Collective conical intersections through light-matter coupling in a cavity}

\date{\today}

\begin{abstract}
   The ultrafast non-radiative relaxation of a molecular ensemble coupled to a
   cavity mode is considered theoretically and by real-time quantum dynamics.
   %
   %
   For equal coupling strength of single molecules
   to the cavity mode, the
   non-radiative relaxation rate from the upper to the lower polariton states
   is found to strongly depend on the number of coupled molecules.
   For $N>2$ molecules, the $N-1$ dark light-matter states between
   the two optically active polaritons feature
   true collective conical intersection crossings, whose location
   depends on the internal atomic coordinates of each molecule in the ensemble,
   and which contribute to the ultrafast non-radiative decay from the upper polariton.
   At least $N=3$ coupled molecules are necessary for cavity-induced collective conical
   intersections to exist
   and, for identical
   molecules, they constitute a special case of the Jahn-Teller effect.
   %
   \end{abstract}


\maketitle


The interaction of atoms
and molecules with quantized light
has the potential to open new routes towards
manipulating their physical and chemical properties, and towards the development of
hybrid matter-light systems with new
attributes~\cite{%
    jay63:89,%
    tav67:714,%
    har89:24,%
    mil05:551,%
    wal06:1325,%
    Asp14:1391%
}.
Over the past few years, ground breaking experiments that realize the
aforementioned scenario using, e.g., microcavities,~\cite{%
    col11:3691,%
    sch11:196405,%
    hut12:1624,%
    sch13:125,%
    ebb16:2403%
    }
have demonstrated the
effective tuning of reaction rates and probabilities~\cite{hut12:1624}, energy
transfer rates among different molecular species~\cite{zho17:9034}
and of molecular vibrations~\cite{geo15:281,geo16:153601}.
A growing body of theoretical
results~\cite{%
    mor07:73001,%
    car12:125424,%
    gal15:41022,%
    geo15:281,%
    sch15:196403,%
    cwi16:33840,%
    gal16:13841,%
    her16:238301,%
    kow16:2050,%
    kow16:54309,%
    fli17:3026,%
    gal17:136001,%
    luk17:4324,%
    fei17:205,%
    ven18:55,%
    sae18:arXiv%
}
has lead, among others, to propose mechanisms to suppress~\cite{gal16:13841} and
catalyze chemical processes~\cite{gal17:136001} through a cavity mode, or
modify the non-adiabatic dynamics of a single molecule strongly coupled
to an electromagnetic mode~\cite{kow16:2050}.

Experimentally, it has been observed that photo-excitation of the upper
polariton branch (UPB) in a coupled cavity-matter system is followed by
population transfer to the lower polariton branch (LPB) before light emission
from the UPB can take place~\cite{col11:3691,sch13:125}.
Time-resolved measurements in hybrid organic dye-molecule systems
indicate that population transfer from the UPB to the
LPB occurs within a time-scale of tens to hundreds of
femtoseconds~\cite{sch13:125}, orders of magnitude shorter than the radiative
life-time of the molecular excited states in isolation.
Theoretical
predictions based on incoherent relaxation rates obtained by Fermi's golden
rule and based on modelling
phonons coupled to the polaritonic excitation also predict relaxation rates
of the order of tens to hundreds of femtoseconds~\cite{agr03:85311,sae18:arXiv}.

Even though it is well established that the ultrafast relaxation rates of molecular
polaritons arise from vibronic interactions of the participant
molecules~\cite{agr03:85311,col11:3691,rib18:6325,sae18:arXiv},
a microscopic, real-time description of such phenomena, which additionally
sheds light onto the connection with standard descriptions of non-radiative
phenomena in chemical systems, is still missing.
In this work, the vibronic interactions leading to ultrafast non-radiative decay
of a molecular ensemble coupled a single electromagnetic mode
will be discussed theoretically and on the basis of
numerically converged quantum dynamics simulations.
It will be shown how
the vibronic origin of the ultrafast relaxation is of similar nature as for
isolated molecular excitations and related to the existence of
collective conical intersections (CCI) among dark states, whose topological properties will
be discussed and compared to their intramolecular counterparts.



 The starting point is an ensemble of non-interacting diatomic molecules
 aligned with the polarization axis of the quantized light mode (also referred
 throughout the paper as
 cavity mode). The Hamiltonian for this system reads
 \begin{align}
     \label{eq:ham1}
     \hat{H} & = \hat{T}_{n} + \hat{H}_\mathsf{el} + \hat{H}_\mathsf{cav} +
      \hat{H}_\mathsf{las},
 \end{align}
 where
 $\hat{T} = \sum_\kappa^N \hat{t}_n^{(\kappa)}$ is the sum of nuclear
 kinetic energy operators for each $\kappa$-th molecule,
 $\hat{H}_\mathsf{el} = \sum_\kappa^N \hat{h}_\mathsf{el}^{(\kappa)}$ is the sum of
 all other intramolecular Hamiltonian terms for each molecule
 \begin{align}
     \label{eq:ham2}
     \hat{h}_\mathsf{el}^{(\kappa)} & =
     \hat{t}_{e}^{(\kappa)} +
     \hat{v}_{ee}^{(\kappa)} +
     \hat{v}_{en}^{(\kappa)} +
     \hat{v}_{nn}^{(\kappa)},
 \end{align}
 $\hat{H}_\mathsf{cav}$ is the cavity and cavity-ensemble Hamiltonian,
 and  $\hat{H}_\mathsf{las}$ describes the eventual coupling to an external
 laser field.
 The terms in Eq.~(\ref{eq:ham2}) correspond to the $\kappa$-th electronic
 kinetic energy and the Coulombic terms represent the electron-electron
 repulsion, electron-nuclei attraction and nuclei-nuclei repulsion,
 respectively.
 The cavity Hamiltonian is given
 by~\cite{Faisal_1987,gal15:41022,fli17:3026,ven18:55}
  \begin{align}
     \label{eq:ham3}
     \hat{H}_\mathsf{cav} & =
     \hbar\omega_c\left(\frac{1}{2}+\hat{a}^\dagger\hat{a}\right) +
     g\; \vec{\epsilon_c}\cdot\vec{\hat{D}}
     \left( \hat{a}^\dagger + \hat{a} \right),
 \end{align}
 where $\omega_c$ is the angular frequency of the cavity mode,
 $\vec{\epsilon}_c$ is its polarization direction, and
 \mbox{$g=\sqrt{\hbar\omega_c/2V\epsilon_0}$} is the coupling strength between the
 cavity and the molecules where $V$ is the quantization volume.
 $\vec{D}=\sum_\kappa^N\vec{\mu}^{(\kappa)}$ is total dipole operator of the
 ensemble.
 %
 In Eq.~(\ref{eq:ham3}), the quadratic dipole self-energy term is being neglected,
 which is only relevant at much higher coupling strengths than considered
 here.
 For further details see, e.g., Refs.~\cite{Faisal_1987,fli17:3026}.
 The coupling to an external laser field is introduced
 semiclassically in the length gauge and dipole approximation as
 $\hat{H}_\mathsf{las} = - \vec{E}(t)\vec{\hat{D}}$, where the electric field
 takes the form $\vec{E}(t)=\vec{\epsilon_L}A(t)\cos(\omega_L t)$.
 It is assumed for simplicity that
 external laser fields couple to the
 molecules and do not pump the cavity mode directly.
 In general this is not necessarily the case and direct coupling to the cavity
 mode can be easily introduced~\cite{sae18:arXiv}. However,
 since strong coupling is assumed (cf. discussion below),
 this model choice has only observable
 consequences at times below the Rabi cycling period of the
 hybrid system, in the order of a few tens of femtoseconds, and is not
 of relevance for our discussion.

 As an illustrative molecular example we consider sodium iodide (NaI), whose
 ultrafast photo-dissociation dynamics in the first excited electronic state
 $^1A$ coupled to the ground state $^1X$ has been the subject of
 extensive experimental and theoretical investigations (See e.g.,
 Ref.~\cite{zew88:1645}), also in the context of cavity-induced
 chemistry~\cite{kow16:2050,ven18:55}.
 Details on the potential energy and transition dipole curves of
 NaI~\cite{ven18:55} and a detailed account of the quantum dynamics numerical
 techniques employed in this work can be found
 elsewhere~\cite{man92:3199,bec00:1,ven18:55}.
 Throughout this work
 the effective cavity-matter coupling is taken as $g/\omega_c$=0.01, where
 $g$ was defined around Eq.~(\ref{eq:ham3}) and can be seen as the rms
 vacuum electric field
 amplitude of the cavity mode~\cite{har89:24}.
 %
 This coupling strength
 is small compared to the single-molecule ultra-strong coupling regime,
 characterized by a Rabi splitting of the polaritonic energy levels (at zero
 detuning) $\hbar\omega_R=2 g\mu_{01}$ comparable to the transition energy.
 The collective Rabi splitting is given by
    \mbox{$\hbar\Omega_R=2 g\mu_{01}\sqrt{N}$}~\cite{tho92:1132},
    where $N$ is the number of coupled molecules and $\mu_{01}$ the transition
    dipole matrix element. For NaI at the Franck-Condon geometry
    this coupling strength results in
    $\hbar\Omega_R=0.13$~eV for $N=1$ and $\hbar\Omega_R=0.30$~eV for $N=5$.
 Rabi splittings of this order have been observed experimentally in
 micro-cavities with coupled organic dye
 molecules~\cite{sch13:125,zho17:9034}.

 %

\begin{figure}[t!]
    \begin{center}
       \includegraphics[width=8.5cm]{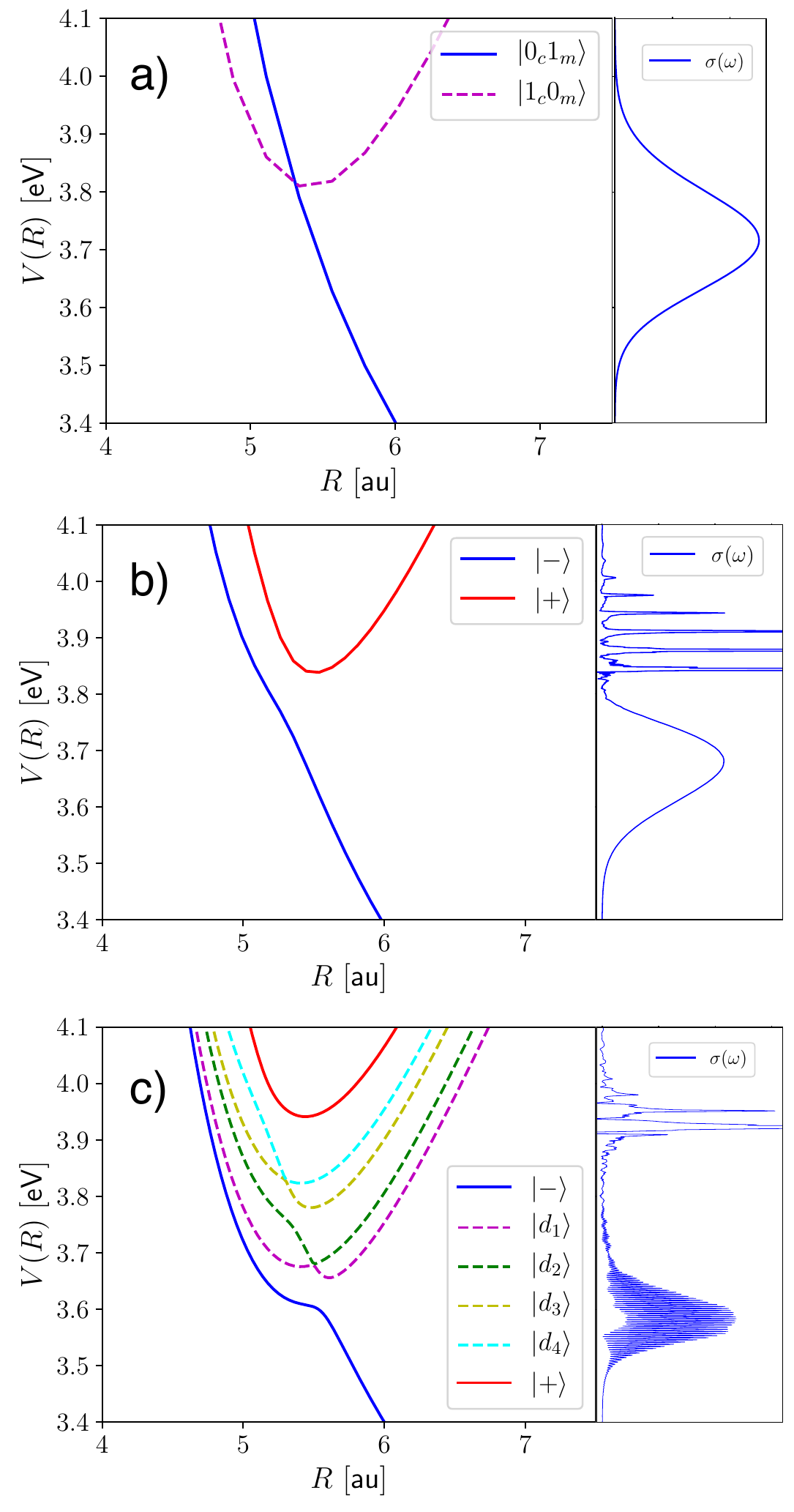}
    \end{center}
    \caption{Left panels: one-dimensional cuts along adiabatic
    polaritonic potential energy surfaces obtained by diagonalization of
    $\mathcal{H}_\mathsf{el-cav}^{[1]}(\mathbf{R})$ (cf. Eq.~(\ref{eq:model})).
    Right panels: absorption spectrum of the hybrid system (blue solid curves)
    and single-molecule photo-dissociation probability at the corresponding
    photon-energy (red dots). Isolated molecule (a), single coupled molecule (b)
    and ensemble of 5 molecules (c). Legends are explained in the main text.
}
    \label{fig:pes_spectra}
 \end{figure}

 We note here that
 for an ensemble of molecules featuring a dissociative excited state
 and coupled to a cavity mode, photo-dissociation can only occur in the LPB.
 This is because, as one of the molecules dissociates, it ceases to be resonant
 with the cavity mode, the lower polariton turns into a pure electronic
 excitation of that molecule and the excitation energy is not available anymore
 for neither the cavity nor the other molecules, which subsequently
 remain in their respective ground
 states (cf. Fig.~(\ref{fig:pes_spectra}c)).
 For this reason, the rate of photo-dissociation directly corresponds to the rate of
 relaxation from the UPB to the LPB, which will be used below.

 We consider first the absorption spectra of a single isolated NaI molecule and
 compare it afterwards to the absorption spectra
 of a molecular ensemble.
 Excitation of an isolated NaI molecule to its first singlet excited electronic
 state results immediately in ballistic dissociation of the nuclear wave packet
 on the corresponding potential energy surface.
 Therefore, the
 photo-absorption spectrum features a single absorption band of,
 in this case, full width at half
 maximum (fwhm) about 0.2~eV centered at an excitation energy of about
 3.7~eV, which is seen in the right panel in
 Fig.~(\ref{fig:pes_spectra}a).
 The absorption spectrum
 \mbox{$\sigma(\omega)\propto\Re\{\omega\int_0^\infty e^{i\omega t} a(t) dt\}$}
 is computed from the
 dipole auto-correlation function
 \mbox{$a(t)=\langle\Psi_\mu(0)|e^{-i\hat{H}t/\hbar}|\Psi_\mu(0)\rangle$}, where
 \mbox{$|\Psi_\mu(0)\rangle=\vec{\epsilon}_L\cdot\vec{\hat{D}}|\Psi_0\rangle$} and
 $|\Psi_0\rangle$ is the ground state of the system.
 The strongly dissociative potential energy surface
 responsible for the fast dissociation is
 shown in the left panel of
 Fig.~(\ref{fig:pes_spectra}a), where the ground state potential
 energy surface shifted by the cavity photon energy
 \mbox{$V_0(R)+\frac{3}{2}\hbar\omega_c$}
 is shown as a dashed curve
 for comparison.
 %

 The photo-dissociation dynamics is significantly changed when the molecules are
 coupled to the cavity mode.
 In the case of a single molecule coupled to the cavity, the absorption spectrum
 is shown in the right panel of Fig.~(\ref{fig:pes_spectra}b). A broad absorption
 band is found in the energy region of the LPB and a set of discrete
 absorption peaks are seen in the region of the UPB. These markedly
 different spectral regions can be explained by the potential
 energy curves of the lower and upper polaritons obtained by diagonalization of
 the $\hat{H}_\mathsf{el}+\hat{H}_\mathsf{cav}$ Hamiltonian as a function of $R$,
 which correspond to dissociative and
 bound potentials, respectively, and are shown in the left panel of
 Fig.~(\ref{fig:pes_spectra}b).

 Population promoted to the LPB immediately dissociates
 as in the isolated molecule case.
 %
 Conversely, the upper polariton features a sharp progression of
 long-lived vibrational excitations, which correspond, to a good approximation, to
 the vibrational energy levels on the ground electronic state potential energy surface
 shifted by the photon energy of the cavity mode.
 Clearly, there is no ballistic photo-dissociation from the upper polariton, as
 the LPB cannot be immediately reached.

 The collective behaviour of a molecular ensemble is now investigated by
 increasing the number of molecules coupled to the cavity mode from $N=1$ to
 $N=5$ while keeping the same single-molecule coupling strength as before. The absorption
 spectrum
 is shown in the right panel in Fig.~(\ref{fig:pes_spectra}c).
 The LPB region features now a broad band with superimposed internal structure.
 The former is indicative of ballistic dissociation along one of the molecular
 coordinates, as in the single molecule case. The latter corresponds to
 bound-type dynamics along the other vibrational degrees of
 freedom~\cite{hel81:368,tannor-book}. These
 dynamics of the LPB are observed independently of the number of molecules in
 the ensemble for $N=2$ up to $N=5$ (not shown) and will not be discussed
 further.
 In the UPB, the linear absorption spectrum still features several peaks
 reminiscent of the vibrational progression of the single molecule case, but
 broader, which is indicative of decay from those states within the first few
 hundreds of femtoseconds after photo-excitation.

The relaxation dynamics of the UPB in real time is accessed by pumping the
system with an external laser of pulse of duration 30~fs (fwhm), photon energy
tuned to the upper polariton region (cf. absorption spectra) and a peak field
amplitude of $5\cdot10^{-4}$~au. This relatively weak field ensures that
in all cases light
absorption takes place in the linear single photon regime.
%
The time-dependent dissociation probability is
defined as
$P_\mathsf{dis}(t)=\langle\Psi(t)|\hat{\Theta}(R_1-R_d)|\Psi(t)\rangle$,
where $\Theta(x)$ is the Heaviside step function and $R_d=15$~au.
 As was mentioned above, the relaxation dynamics from the upper to the lower
 polariton branches can be traced through the probability of photo-dissociation,
 which can only occur in the LPB.

\begin{figure}[t!]
    \begin{center}
       \includegraphics[width=8.5cm]{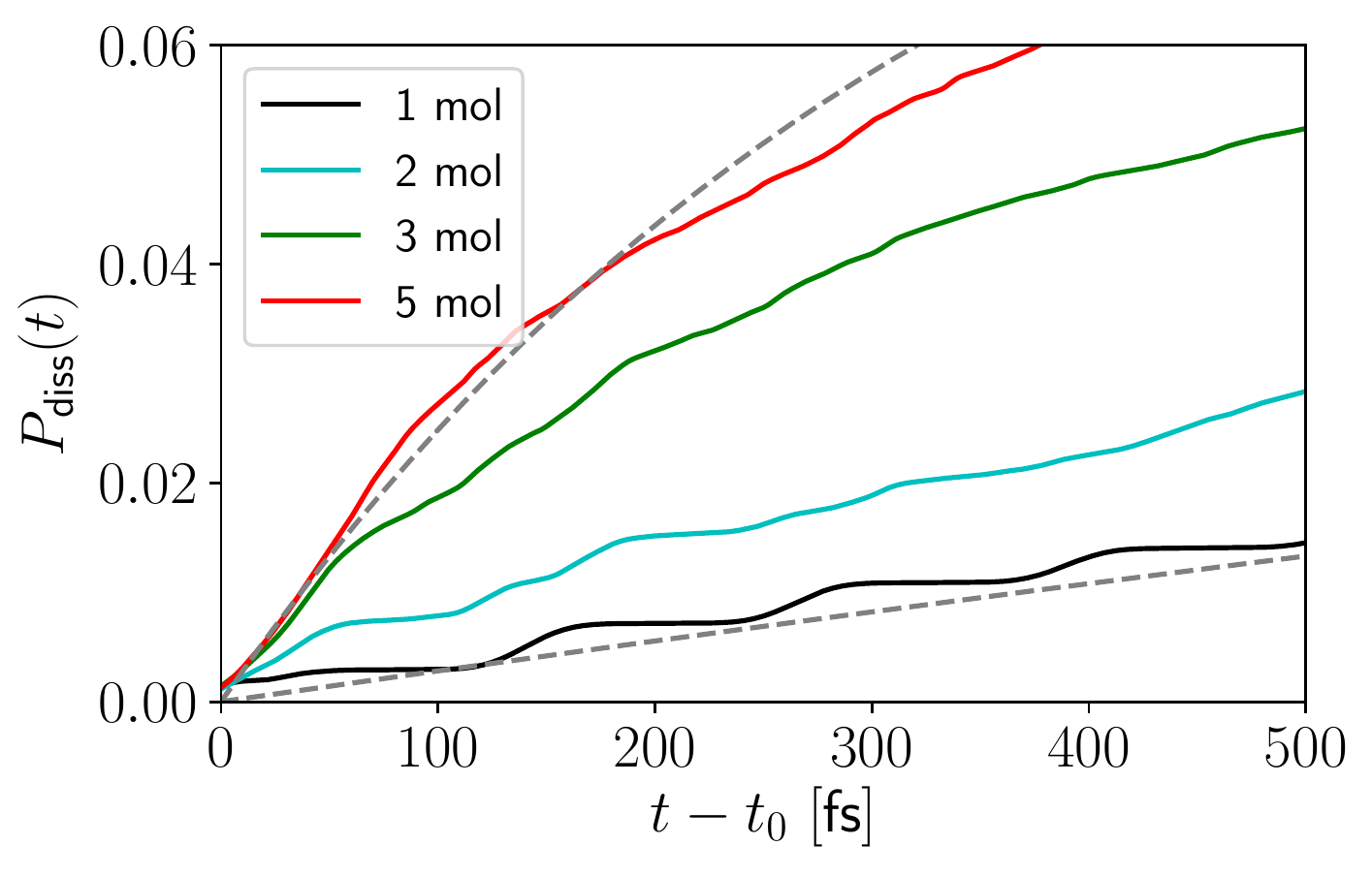}
    \end{center}
    \caption{Single-molecule photo-dissociation probability
        $P_\mathsf{dis}(t)=\langle\Psi(t)|\Theta(R_1-R_b)|\Psi(t)\rangle$
        as a function of time for an ensemble of different size
        interacting with a
        laser pulse of
        photon energy
        $\hbar\omega_L=3.9$~eV. $R_b=$~15~au. The dashed curves correspond to
        fitted first order rate expressions.}
    \label{fig:diss_prob}
 \end{figure}

 Inspection of $P_\mathsf{dis}(t)$ in Fig~(\ref{fig:diss_prob}) for different
 ensemble sizes pumped to the UPB indicates that the
 dissociation occurs progressively, as compared to ballistic
 dynamics in the LPB.
 Even though the individual
 coupling of each molecule to the cavity, the laser parameters, and the total
 amount of population pumped by the laser to the UPB are the same
 in all cases, the rate of dissociation and therefore the rate of relaxation
 significantly depends on the size of the ensemble.

 The calculated population curves in Fig~(\ref{fig:diss_prob}) can be used to
 fit effective
 first order relaxation rate constants $\kappa_N$ to the LPB starting from the UPB.
 For $N=5$, $\kappa_5^{-1}\approx 350$~fs, whereas
 $\kappa_1^{-1}\approx 3800$~fs. The first order rates result in the
 superimposed dashed
 curves in Fig.~(\ref{fig:diss_prob}),
 which however do not completely capture all features of the quantum
 dynamics at short times.
 This marked dependence of the decay rate on the number of molecules is not found
 in expressions based on Fermi's golden rule and an assumed density
 of vibrational states~\cite{agr03:85311,sae18:arXiv}, and
 hints at the participation of
 decay pathways involving non-adiabatic nuclear dynamics with coupling between the
 electronic-polaritonic states and the vibrational degrees of freedom.



 In order to shed light onto the nature of the relaxation pathways and to
 understand the origin of these different kinds of dissociative
 dynamics, we direct our attention to the dark states found between the
 upper and lower polaritons.
 In the Tavis-Cummings model of an ensemble of
 two-level atoms coupled to a cavity, and in the zero-detuning case, all dark
 states are degenerate, appear at the average energy of the two bright
 polaritons and feature no dipole coupling to the ground state of the
 hybrid system~\cite{tav67:714}.
 Their dark nature is clearly manifest in Fig.~(\ref{fig:pes_spectra}c) by the
 practical lack of absorption between $3.65$ and $3.85$~eV in the linear
 absorption spectrum.
 When nuclear motion is present, the degeneracy of the dark states can
 be lifted through nuclear displacements that modulate the energy gap of the
 corresponding molecule. This leads to CCI among the dark
 polaritonic states, which are
 referred to as \emph{collective} to emphasize the fact that their location in
 coordinate space depends on the internal coordinates of the different molecules
 in the ensemble.
 The existence of points of intersection among dark polaritonic states has been
 noted recently as well by Feist and collaborators~\cite{fei17:205}.

 A cut through the potential energy surfaces obtained by diagonalization of the
 $\hat{H}_\mathsf{el}+\hat{H}_\mathsf{cav}$ Hamiltonian in the single-excitation
 space (SES) and for $N=5$ is shown in the left panel of
 Fig.~(\ref{fig:pes_spectra}c).
 For five molecules, six polaritonic states are present in the SES, two of which
 are the bright polaritons and four of them are nominally dark. In the cut
 shown, $R_2=R_3=5.35$~au, $R_4=R_5=5.5$~au, and $R_1$ is scanned between $4$
 and $7.5$~au. Two CCI, the origin of which will be
 discussed below, are seen along this PES cut at precisely
 $R_1=5.35$ and $R_1=5.5$, which is not fortuitous.

%
 The matrix representation of the $\hat{H}_{\mathsf{el}} + \hat{H}_{\mathsf{cav}}$
 operator in the basis of non-interacting cavity-ensemble states and in the SES
 results in
 the molecular Tavis-Cummings Hamiltonian~\cite{support}
 \begin{align}
     \label{eq:model}
     \mathcal{H}_\mathsf{el-cav}^{[1]}=
 \begin{pmatrix}
     \hbar\omega_c     & \gamma^{(1)}(R_1)  &  \gamma^{(2)}(R_2) & \gamma^{(3)}(R_3)  & \cdots \\
    \gamma^{(1)}(R_1)  & \Delta^{(1)}(R_1)  &  0                 & 0                  & \cdots \\
    \gamma^{(2)}(R_2)  & 0                  &  \Delta^{(2)}(R_2) & 0                  & \cdots \\
    \gamma^{(3)}(R_3)  & 0                  &  0                 & \Delta^{(3)}(R_3)  & \cdots \\
    \vdots             & \vdots             &  \vdots            & \vdots             & \ddots \\
  \end{pmatrix},
 \end{align}
 %
 where $\Delta^{(\kappa)}(R_\kappa)=V_1^{(\kappa)}-V_0^{(\kappa)}$ is the energy
 gap of the $\kappa$-th molecule and
 $\gamma^{(\kappa)}(R_\kappa)=g\mu_{01}^{(\kappa)}$ is the dipole
 coupling of the $\kappa$-th molecule to the cavity mode.
 Hamiltonian~(\ref{eq:model}) has the form of an arrowhead matrix,
 whose properties have been investigated in
 the contexts of applied mathematics~\cite{wil65:algebra,oe90:497}
 and molecular physics~\cite{wal84:729}.
 %

 %
 The most important property of arrowhead matrices for our purposes is the fact
 that for every $m$ molecules with the same energy gap $\Delta$, there is an
 eigenvalue \mbox{$\lambda_j=\Delta$} of multiplicity $m-1$~\cite{oe90:497}.
 Hence, for $m=2$ molecules with the same energy gap, there is one eigenstate
 of $\mathcal{H}_\mathsf{el-cav}^{[1]}$
 at the corresponding energy value. In case $m=3$ molecules have the same
 energy gap, e.g., molecules $1$ to $3$,
 there are two degenerate eigenstates states at that energy resulting in a CCI
 of order two.
 For the case of identical molecules, this degeneracy is found in the
 one-dimensional
 space \mbox{$R_1=R_2=R_3$} and it is lifted by displacements in the two-dimensional branching
 space that removes this equality. 
 The local symmetry of the identical molecules giving rise to the CCI can be
 described in the permutation symmetry group $S_3$~\cite{bunker-book:symmetry},
 which among others is isomorph with the $D_3$ point group of an equilateral
 triangle~\cite{koep10}.
 As is well known, molecules of $D_3$ and related symmetry point groups, e.g. $C_{3v}$, are
 Jahn-Teller active~\cite{dom04,koep10} and the degeneracy of electronic state
 energies is lifted linearly for displacements
 out of the highly symmetric atomic arrangement, resulting in
 conically intersecting potential energy
 surfaces.
 Therefore, cavity-induced CCI are, from a mathematical standpoint,
 analogous to the commonly encountered~\cite{Tru03:32501}
 intra-molecular conical intersection case~\cite{%
     yar96:985,%
     yar98:511,%
     wor04:127,%
     dom04,%
     koep10%
 },
 including their divergent non-adiabatic coupling
 matrix elements
 \mbox{$\langle \psi_i|\vec{\nabla}_R |\psi_j\rangle$}
 at the regions of intersection~\cite{support}, where
 $|\psi_j\rangle$ are the eigenstates of $\mathcal{H}_\mathsf{el-cav}^{[1]}$.

 Generalising to larger numbers of molecules, the molecular Tavis-Cummings
 Hamiltonian features a CCI of order $m-1$ for any subset of $m$ molecules at
 molecular geometries that result in an equal energy gap for all molecules of
 the subset.
 As a final example, the polaritonic states resulting of four identical molecules with the same
 geometry can be classified according to the symmetry representations of the
 $S_4$ group, which is isomorph with the $T_d$ point group of the tetrahedron
 and which leads to triply degenerate potential energy crossings and to the $T_2
 \otimes t_2$ Jahn-Teller effect~\cite{koep10}.
 %



 Summarizing, real-time wave packet simulations of a molecular ensemble coupled
 to a cavity mode show that the non-radiative energy relaxation rate from
 the upper to the lower polaritonic states is strongly dependent on the number of
 coupled molecules.
 %
 %
 This is in contrast to descriptions based on Fermi's golden rule rates.
 Once the system reaches the manifold of nominally dark states, the vibronic
 relaxation dynamics proceeds through cavity-induced CCI, which mathematically
 are a consequence of the arrowhead-matrix form of the molecular
 Tavis-Cummings Hamiltonian in the SES.
 From a mechanistic perspective, the conical
 intersection topology funnels nuclear wave packets through it by first
 attracting them when in the upper part of the cone and then pushing them away
 once the probability amplitude appears on the lower electronic
 state~\cite{dom04}. This mechanistic idea
 represents the cornerstone of ultrafast non-radiative relaxation
 in isolated molecular
 systems.
 In the context of polariton relaxation, CCI may
 be an important ingredient for fast localization and decay of
 collectively coupled excitations.
 Since only the energy gap is determinant for the existence of the
 CCI~\cite{oe90:497}, these are expected to be robust against molecular
 rotations or other external perturbations that modulate the off-diagonal coupling
 strength of individual molecules to the cavity mode, as well as to local
 interactions with an environment, which will lead only to displacements of the
 locus of intersection.
 The precise role of CCI in collective decay mechanisms, specially for larger
 ensembles and more complex molecules,
 remains to be further investigated.
 %
 %
 %

\begin{acknowledgments}
    I want to thank L.B. Madsen, L.S. Cederbaum, H.-D. Meyer and J. Feist
    for insightful discussions.
\end{acknowledgments}

\bibliography{cites}






\end{document}